\documentstyle[bo99,epsfig]{article}

\def\gapp{\ifmmode\stackrel{>}{_{\sim}}\else$\stackrel{>}{_{\sim}}$\fi}

\title{Discovery of two high-magnetic-field \\ radio pulsars}

\author{V. M. Kaspi $^1$, F. Camilo $^2$, A. G. Lyne $^2$, R. N. Manchester
$^3$, J. F. Bell $^3$, N. D'Amico $^4$, N. P. F. McKay $^2$, F. Crawford $^1$ }

\affil{
1) Department of Physics and Center for Space Research, 70 Vassar Street,
MIT, Cambridge, MA 02139\\
2) University of Manchester, Jodrell Bank Observatory, Macclesfield, 
Cheshire, SK11~9DL, UK\\
3) Australia Telescope National Facility,
  CSIRO, P.O.~Box~76, Epping, NSW~1710, Australia\\
4) Osservatorio Astronomico di Bologna, via Ranzani 1, 40127 Bologna and
Istituto di Radioastronomia del CNR, via Gobetti 101, 40129 Bologna, Italy}

\begin{document}

\maketitle

\begin{abstract}

We report the discovery of two isolated radio pulsars having the
largest inferred surface dipole magnetic fields yet seen in the population:
$4.1\times 10^{13}$\,G and $5.5\times 10^{13}$\,G.  These pulsars
show apparently normal radio emission in a regime of magnetic field
strength where some models predict no emission should occur.  They
have spin parameters and magnetic fields similar to those of some
magnetar candidates, but exhibit very different radiative properties.
This demonstrates that if the putative magnetars are indeed isolated
neutron stars, their unusual attributes cannot be only a consequence
of their large inferred magnetic fields.

\end{abstract}

\section{ Introduction }

Recently it has been suggested (Thompson \& Duncan 1992, Kulkarni \&
Frail 1993, Vasisht \& Gotthelf 1997,
Kouveliotou et al. 1998) that, in addition to the radio pulsars, there
exists a class of isolated rotating neutron stars with ultra-strong
magnetic fields -- the so-called ``magnetars.''  The observational
properties of radio pulsars and putative magnetars are very different.
Known radio pulsars, whose spin periods span the range from 
0.0015--8.5~s, rarely have observable X-ray pulsations, and, in all
cases, the X-ray power is much smaller than the spin-down luminosity.
By contrast, the objects that have been suggested to be magnetars,
namely soft gamma repeaters (SGRs) and anomalous X-ray pulsars (AXPs),
exhibit pulsations with periods 5--12~s, and high energy emission that
is many orders of magnitude stronger than their spin-down luminosity
(Mereghetti \& Stella 1995).  Their pulsations have gone undetected at
radio wavelengths.  The dichotomy is thought to be a result of the
much larger magnetic fields in magnetars, with magnetic field decay
heating the neutron star to produce thermal X-ray emission (Thompson
\& Duncan 1993), or thermal emission from initial cooling enhanced
by the large field (Heyl \& Hernquist 1997).

Here we report the discovery of two isolated radio pulsars, PSRs
J1119$-$6127 and J1814$-$1744, which have the largest inferred surface
magnetic fields yet seen among radio pulsars.  The results reported
here will be described in more detail by Camilo et al. (2000).

\section{ Observations and Results }

PSRs~J1119$-$6127 and J1814$-$1744 were discovered as part of an
ongoing survey of the Galactic Plane using the 64-m Parkes radio
telescope (Lyne et al. 2000, see also D'Amico et al., these
proceedings).  PSR~J1119$-$6127 has $P = 0.41$\,s and period
derivative $\dot{P} = 4.0\times 10^{-12}$, the largest known among
radio pulsars.  We have also measured an apparently stationary period
second derivative, $\ddot{P} = -4\times 10^{-23}$~s$^{-1}$, making
this only the third pulsar for which this has been possible through
absolute pulse numbering.  PSR~J1814$-$1744 has $P=4.0$\,s and
$\dot{P}=7.4 \times 10^{-13}$.  Spin and astrometric parameters for
both pulsars are given in Table~1.  From the standard equation for the
dipolar surface magnetic field $B = 3.2 \times 10^{19} (P
\dot{P})^{1/2}$~G, we infer surface magnetic fields of $4.1 \times
10^{13}$\,G and $5.5 \times 10^{13}$\,G for PSRs~J1119$-$6127 and
J1814$-$1744, respectively.  These are the highest magnetic field
strengths yet observed among radio pulsars.

\section{ Discussion }

Figure~1 is a plot of $\dot{P}$ versus $P$ for the radio pulsar
population, with PSRs~J1119$-$6127 and J1814$-$1744 indicated.  Also
shown are the sources usually identified as magnetars,
namely the five AXPs and two SGRs for which $P$ and $\dot{P}$ have
been measured.

\begin{table}[t]
\begin{center}
\begin{tabular}{lll}
\multicolumn{3}{c}{Table~1: Parameters for pulsars J1119$-$6127 and
  J1814$-$1744.} \\
\hline\hline
Right ascension (J2000)                 & $11^{\rm h}19^{\rm m}14^{\rm s}.2(1)$
                                          & $18^{\rm h}14^{\rm m}43^{\rm s}.0(2)
$                                                                             \\
Declination (J2000)                     & $-61^\circ 27' 48''.3(6)$
                                          & $-17^\circ 44' 47(23)''$          \\
Spin period, $P$ (s)                    & 0.4076034160(1) & 3.975823037(1)    \\
Period derivative, $\dot P$             & $4.022930(4)\times 10^{-12}$
                                          & $7.434(4)\times 10^{-13}$         \\
Period second derivative, $\ddot P$ (s$^{-1}$) &$-4.1(3)\times10^{-23}$ &\dots\\
Period epoch (MJD)                      & 51075.0 & 51075.0                   \\
Dispersion measure (cm$^{-3}$\,pc)      & 713(20) & 834(20)                   \\
Flux density at 1374\,MHz, $S$ (mJy)    & 0.7(2) & 0.5(2)                     \\
Surface magnetic field, $B$ (Gauss)     & $4.1\times 10^{13}$
                                        & $5.5\times 10^{13}$                 \\
Characteristic age, $P/(2\dot P)$ (kyr) & 1.6   & 85                          \\
Spin-down luminosity (erg\,s$^{-1}$)    & $2.3\times10^{36}$
                                          & $4.7\times10^{32}$                \\
Distance, $d$ (kpc)         & 2.4--8  & $10(2)$                   \\
Radio luminosity at 1374\,MHz, $S d^2$ (mJy\,kpc$^2$) & $\sim 20$ & $\sim 50$ \\
Braking index, $n$                      & 3.0(1)  & \dots                     \\
\hline
\end{tabular}
\end{center}
\end{table}
          
\begin{figure}[thb]
\centerline{\psfig{file=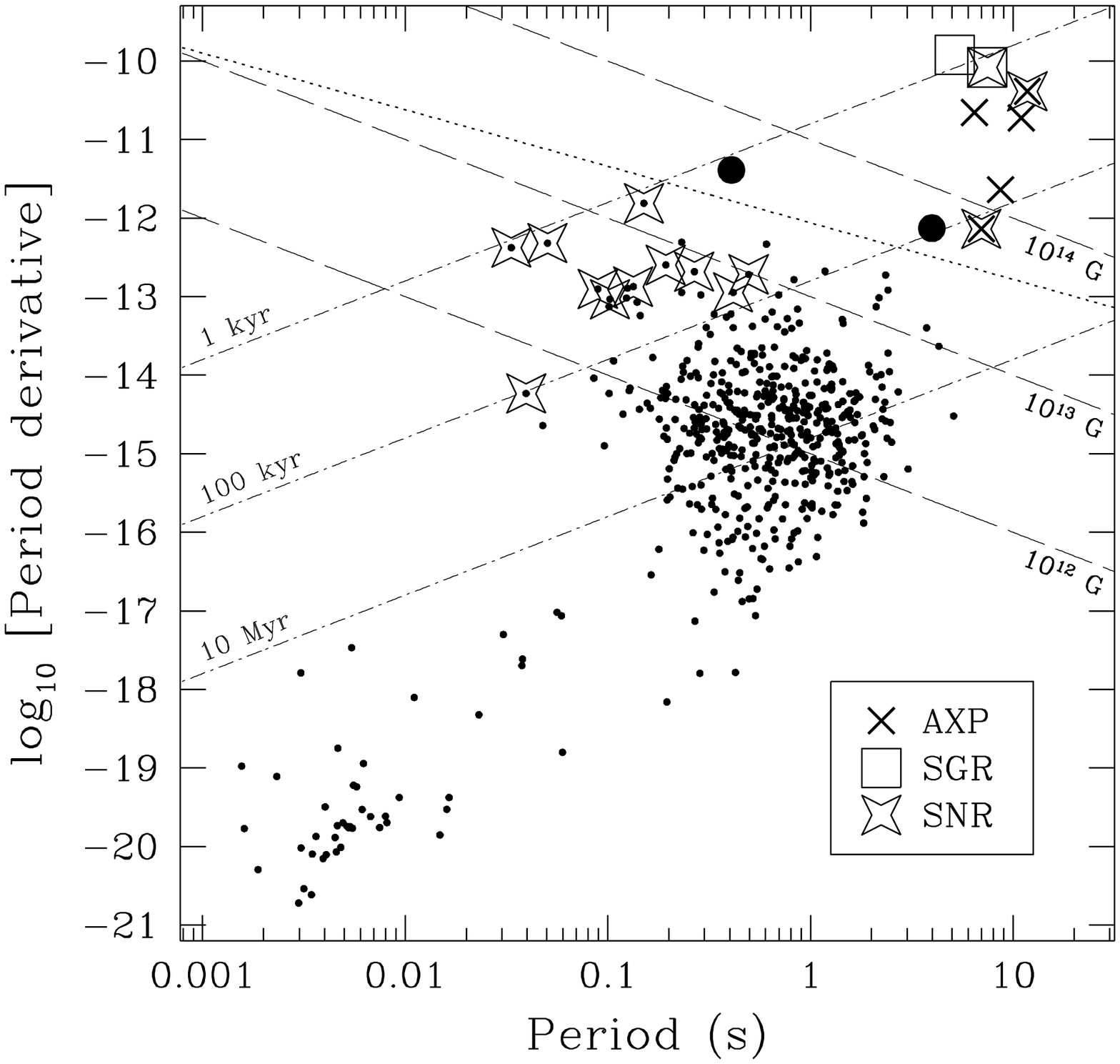,height=7cm}}
\caption[]{$P-\dot{P}$ diagram for radio pulsars, with SGRs and AXPs
indicated.  PSRs J1119$-$6127 and J1814$-$1744 are indicated with large
solid circles.  Lines of constant magnetic field are shown dashed, and
lines of constant characteristic age ($\tau_c \equiv P/2\dot{P}$)
are shown dot-dashed.  The dotted line is the proposed illustrative
radio-loud/radio-quiet boundary (Baring \& Harding 1998). Radio pulsars plausibly
associated with supernova remnants (SNRs) are indicated with four-point stars.}
\end{figure}

Most models of the radio emission physics depend on pair-production
cascades above the magnetic poles and hence on the strength of the
magnetic field.  However, at field strengths near or above the quantum
critical field, $B_c \equiv {m_e^2 c^3}/{e \hbar} = 4.4 \times
10^{13}$~G, the field at which the cyclotron energy is equal to the
electron rest-mass energy, processes such as photon splitting may
inhibit pair-producing cascades. It has therefore been argued (Baring
\& Harding 1998) that a radio-loud/radio-quiet boundary can be drawn
on the $P$--$\dot{P}$ diagram, with radio pulsars on one side, and
AXPs and SGRs on the other (see Fig.~1).  The existence of
PSRs~J1119$-$6127 and J1814$-$1744, however, demonstrates that radio
emission can be produced in neutron stars with surface magnetic fields
equal to or greater than $B_c$.

Especially noteworthy is the proximity of PSR~J1814$-$1744 to the
cluster of AXPs and SGRs at the upper right corner of Figure~1.  In
particular, this pulsar has a nearly identical $\dot{P}$ to the AXP
1E~2259+586 (Fahlman \& Gregory 1981, Baykal et al. 1996, Kaspi et
al. 1999).  The disparity in their emission properties is therefore
surprising.  The radio emission upper limit (Coe et al. 1994) for
1E~2259+586 implies a radio luminosity at 1400\,MHz of
$<$0.8\,mJy\,kpc$^2$.  This limit is comparable to the lowest values
known for the radio pulsar population.  That the radio pulse may be
unobservable because of beaming cannot of course be ruled out.

The radio-loud/radio-quiet boundary line displayed in Figure~1 is more
illustrative than quantitative.  However, the apparently normal radio
emission from PSRs~J1119$-$6127 and J1814$-$1744, and the absence of
radio emission from AXP 1E~2259+586, suggests that it will be
difficult to delineate any such boundary without fine model-tuning.
Furthermore, Pivovaroff et al.  (2000)
show, from archival data, that PSR~J1814$-$1744 must be
significantly less X-ray luminous than 1E~2259+586.  The similar spin
parameters for these two stars, and in turn the common features
between 1E~2259+586 and the other AXPs and SGRs, imply that very high
inferred magnetic field strengths cannot be the primary factor
governing whether an isolated neutron star is a magnetar.

PSR~J1119$-$6127 is notably young.  Only three other pulsars having
ages under 2\, kyr are known: the Crab pulsar ($\tau = 1.3$\,kyr),
PSR~B1509$-$58 ($\tau=1.6$\,kyr), and PSR~B0540$-$69
($\tau=1.7$\,kyr).  The age of a pulsar is given by $\tau = [ 1- (
{P_0}/{P})^{n-1}](P/(n-1)\dot{P}) \simeq P/2\dot{P}$, where
$P_0$ is the spin period at birth (generally assumed to be much
smaller than the current spin period) and $n$ is the ``braking
index,'' defined via the relation for the spin evolution $\dot{\nu}
\propto \nu^n$, where $\nu \equiv 1/P$.  In the standard oblique
rotating vacuum dipole model, $n \equiv \nu \ddot{\nu}/(\dot{\nu})^2 =
3$.  For PSR~J1119$-$6127, the parameters listed in Table~1 imply
$n=3.0\pm0.1$.  This is the first measured braking index for a pulsar
that is consistent with the usually assumed $n = 3$.

\end{document}